\documentclass[12pt]{article}

\usepackage{graphicx,amssymb,url,mathrsfs, amsmath}
\usepackage{eucal}
\usepackage{wrapfig}
\usepackage{boxedminipage}
\usepackage{setspace}
\usepackage{subfigure}
\usepackage[dvipdfm]{hyperref}
\usepackage{amsfonts}

\newcommand{\arctanh}[1]{\text{arctanh}}

\begin{document}

\begin{titlepage}

\begin{flushright}
{\tt hep-th/...}
\tt {FileName:....tex} \\
{\tt \today}

\end{flushright}
\vspace{0.5in}

\begin{center}
{\large \bf  Diffractive and deeply virtual Compton scattering in holographic QCD}\\
\vspace{10mm}
Alexander Stoffers and Ismail Zahed\\
\vspace{5mm}
{\it \small Department of Physics and Astronomy, Stony Brook University, Stony Brook, NY 11794, USA}\\
          \vspace{10mm}
{\tt October 13, 2012}
\end{center}
\begin{abstract}

We further analyze the holographic dipole-dipole scattering amplitude developed in \cite{Basar:2012jb,Stoffers:2012zw}. 
Gribov diffusion at strong coupling yields the scattering amplitude in a confining background. We compare the holographic result for the differential cross section to proton-proton and deeply virtual Compton scattering data. 
 
\end{abstract}
\end{titlepage}

\renewcommand{\thefootnote}{\arabic{footnote}}
\setcounter{footnote}{0}



\section{Introduction}

In \cite{Basar:2012jb,Stoffers:2012zw} a holographic version of the dipole-dipole scattering approach \cite{Mueller:1989st, Mueller:1994gb,Mueller:1993rr,Mueller:1994jq,Iancu:2003uh, Nikolaev:1990ja,Nikolaev:1991et,Salam:1995zd,Salam:1995uy,Navelet:1996jx,Navelet:1997tx,GolecBiernat:1998js} in the Regge limit is used to describe high energy hadron-hadron scattering within the context of holographic QCD. In this limit, the scattering amplitude is dominated by pomeron exchange, i.e. exchange of ordered gluon ladders with vacuum quantum number. The holographic pomeron is argued \cite{Basar:2012jb,Stoffers:2012zw} to be the exchange of a non-critical, closed string in transverse AdS$_3$.  

Using the dipole-dipole scattering approach, two Wilson loops are correlated via a minimal surface with string tension $\sigma_T$. In the presence of a large rapidity gap $\chi$ and large impact parameter $b$, the closed string exchange is T-dual to an open string exchange subjected to a longitudinal 'electric' field $E=\sigma_T\,{\rm tanh}(\chi/2)$ that causes the oppositely charged string end-points to accelerate~\cite{Basar:2012jb}. This acceleration induces an Unruh temperature $T_U\approx \chi/2\pi b$ in the middle of the string worl-sheet. For large impact parameter, the Unruh temperature is low and only the tachyon mode of the non-critical string is excited. This tachyonic string mode is diffusive in curved AdS$_3$, which is reminiscent of Gribov's diffusion in QCD. In particular, the properly normalized diffusion kernel with suitable boundary conditions in the infrared yields a {\it wee-dipoles} density that is similar to the QCD one in the conformal limit.  The convolution of the two {\it wee-dipole} densities yields the eikonalized scattering amplitude and allows for a 'partonic' picture similar to \cite{Brodsky:2006uqa,Brodsky:2006uq}, albeit at strong coupling.

The holographic, strong coupling description allows to access the saturation regime at small Bjorken $x$ and small momentum transfer. In \cite{Stoffers:2012zw} the holographic dipole-dipole cross section was compared to DIS data from HERA. This leads to a fit of the t'Hooft coupling $\lambda$ through the slope of the proton structure function $F_2$, while the remaining parameters were adjusted to be in reasonable agreement with QCD expectations. The scattering amplitude was obtained both in a confining background with a hard wall, as well as in the conformal limit. Both backgrounds yield a cross section that is comparable to the data. 

Exclusive diffractive processes such as proton-proton ($pp$) diffraction and deeply virtual Compton scattering (DVCS) reveal information about the proton shape in the transverse plane. We will use the holographic dipole-dipole scattering model, \cite{Basar:2012jb,Stoffers:2012zw}, to describe $pp$ diffraction and DVCS. At large momentum transfer, the holographic differential $pp$ cross section is sensitive to length scales of the typical string length or $1/\sqrt{\sigma_T}$. In the confining background it is of the order of the IR cutoff. To get a better fit on the two parameters concerning the effective size of the proton and the IR cutoff, we compare the holographic result for diffractive proton-proton and DVCS cross sections with the data.

The dipole picture has been used to describe exclusive diffractive processes, see for example \cite{Shoshi:2002in}, \cite{Kowalski:2006hc}, 
\cite{Donnachie:2000px,McDermott:2001pt, Favart:2004uv,Kopeliovich:2008ct}.
Within the gauge/gravity duality, hadron-hadron scattering and the holographic pomeron has been discussed in numerous places, see e.g. \cite{Rho:1999jm,Basar:2012jb,Stoffers:2012zw,Janik:1999zk,Janik:2001sc,Janik:2000aj, Janik:2000pp, Polchinski:2001tt, Polchinski:2002jw, Brower:2006ea,Brower:2007xg, Andreev:2004sy,Andreev:2004jm, Hatta:2007cs, Hatta:2007he,Albacete:2008ze, Albacete:2008vv,
Cornalba:2008sp,Cornalba:2010vk}. In particular, $pp$ diffraction was studied in \cite{Domokos:2009hm,Domokos:2010ma} and DVCS in \cite{Gao:2009se,Marquet:2010sf,Nishio:2011xz, Costa:2012fw}.  Our construction relies on the holographic results established in~\cite{Basar:2012jb,Stoffers:2012zw} which involve a non-critical string in $D_\perp=3$ dimensions and whereby the pomeron is the tachyon mode at large impact parameter.
They are rooted in the widely used approach to dipole-dipole scattering at high-energy.
This paper is structured as follows. We will briefly review the main results of \cite{Basar:2012jb,Stoffers:2012zw} in section 2 before comparing the differential $pp$ cross section to CERN ISR and LHC data in section 3. A comparison to DVCS data from HERA is done in section 4 and the conclusions given in section 5.

\section{Diffractive scattering as dipole-dipole scattering}

In the dipole-dipole scattering approach at high energies ($\chi = \ln (\frac{s}{s_0})\gg 1$), the scattering amplitude for the process $a \ p \rightarrow c \ p$ factorizes and can be written as
\begin{eqnarray}
{\cal T}(\chi, {\bf b}_\perp) = \int_0^{\infty}  du \int_0^{\infty}  du' \  \psi^*_{a}(u)  \psi^*_p(u') \ {\mathcal T}_{DD}(\chi, {\bf b}_\perp, u, u') \ \psi_{b}(u)  \psi_p(u'), \label{amplitude}
\end{eqnarray}
with transverse impact parameter ${\bf b}_\perp$, rapidity $\chi$ and dipole-dipole amplitude ${\mathcal T}_{DD}$. The wave functions $\psi$ are parametrized by $u=-{\rm ln}(z/z_0)$, with the effective dipole size $z$ and the IR cutoff $z_0$.  The dipole-dipole amplitude is evaluated using the gauge/gravity duality and the virtuality of the scatterers is identified with the holographic direction of the curved space \cite{Stoffers:2012zw}, \cite{Polchinski:2001tt,Polchinski:2002jw},  \cite{Brodsky:2006uqa,Brodsky:2006uq}.

In the eikonal approximation the differential cross section reads
\begin{eqnarray}
\frac{d \sigma_{{\tiny{a p \rightarrow c p}}}}{dt}(\chi, |t|) &=& \frac{1}{16 \pi s^2} |{\cal T} (\chi, |t| )|^2 \label{amplitude} \\
&=&\frac{1}{4 \pi} \left|i \int  d{\bf b}_\perp \int  du \int  du' \  e^{iq_\perp \cdot {\bf b}_\perp} \ |\psi_{ab} (u)|^2 |\psi_{p} (u')|^2 \ (1-e^{{\bf WW}}) \right|^2  \label{dsigmadt}\\
&=& \frac{\pi}{4} \left| i \int \ d|{\bf b}_\perp|^2 \int  du \int du' \ J_0(\sqrt{|{\bf b}_\perp|^2|t|}) \ |\psi_{ab} (u)|^2 |\psi_p (u')|^2 \ (1-e^{{\bf WW}}) \right|^2 \nonumber \\ \label{dsigmadtbessel} 
\end{eqnarray}
with $t=- q_\perp^2$. Here, $J_0$ is the Bessel function and the overlap amplitude is defined by $|\psi_{ab} (u)|^2 \equiv \psi^*_a(u)\psi_b(u)$. Note that the scattering amplitude in (\ref{amplitude}) is purely imaginary.
In \cite{Basar:2012jb,Stoffers:2012zw} the eikonal ${\bf WW}$, which is the correlator of two Wilson loops, was obtained through closed string exchange in a weakly curved, confining space.  The string exchange can be viewed as a funnel connecting the two dipoles at a holographic depth $z$ and $z'$. Identifying these positions $z$, $z'$ of the endpoints of the funnel with the effective size of the dipoles gives rise to a density ${\bf N}$ of {\it wee-dipoles} surrounding each parent dipole. For $D_\perp=3$, we identify \cite{Stoffers:2012zw}
\begin{eqnarray}
{\bf WW} \approx - \frac{g_s^2}{4} \left(2\pi \alpha' \right)^{3/2} z z' \ {\bf N}(\chi, z,z', {\bf b}_\perp)  \ . \label{WW}
\end{eqnarray} 
We consider transverse AdS$_3$
\begin{eqnarray}
ds_\perp^2=\frac{1}{z^2}(d{\bf b}_\perp^2+dz^2) \ , \label{metric}
\end{eqnarray}
with a cutoff (hard wall) imposed at some $z_0$. The {\it effective} string tension will be defined as
\begin{eqnarray}
g_s\equiv \kappa \frac{1}{{4\pi\alpha'}^2 N_c}\equiv\kappa \frac{{\lambda}}{4\pi N_c} \ ,
\label{GS}
\end{eqnarray}
with t'Hooft coupling $\lambda$ and $\alpha'=1/(2\pi\sigma_T)\equiv 1/\sqrt{\lambda}$ the string tension (in units of the AdS radius). $N_c$ is the number of colors and the parameter $\kappa$ is fixed by the saturation scale, see \cite{Stoffers:2012zw}.
The density reads 
\begin{eqnarray}
{\bf N}(\chi, {\bf b}_\perp, z, z')=\frac{1}{zz'}\,\Delta(\chi,\xi)+\frac{z}{z'z_0^2}\,\Delta(\chi,\xi_*) \ , \label{dipoledensity}
\end{eqnarray}
and the heat kernel $\Delta$ in the background (\ref{metric}) is given by
\begin{eqnarray}
\Delta_\perp(\chi,\xi)=\frac{e^{(\alpha_{\bf P}-1) \chi}}{(4 \pi {\bf D} \chi)^{3/2}} \frac{\xi e^{-\frac{\xi^2}{4{\bf D} \chi}}}{\sinh(\xi)} \ ,
\end{eqnarray}
with the chordal distances ($u=-{\ln}(z/z_0)$)
\begin{eqnarray}
{\rm cosh}\xi&=&{\rm cosh}(u'-u)+\frac 12 {\bf b}_\perp^2\,e^{u'+u}\, \nonumber \\
{\rm cosh}\xi_*&=&{\rm cosh}(u'+u)+\frac 12 {\bf b}_\perp^2 e^{u'-u}\, \ .
\end{eqnarray}
To leading order in $1/\sqrt{\lambda}$ the pomeron intercept and the diffusion constant read
\begin{eqnarray}
\alpha_{\bf P}&=&1+\frac{D_\perp}{12} \nonumber  \ , \\
{\bf D}&=& \frac{\alpha'}2=\frac 1{2\sqrt{\lambda}} \ .
\label{HOLO2}
\end{eqnarray}

In order to confront the holographic result for the differential proton-proton cross section with the data, we have to fix the parameters entering the eikonal ${\bf WW}$ in (\ref{WW}). In \cite{Stoffers:2012zw} a comparison of the proton structure function $F_2$ to DIS data determined the following numerical values. $N_c$ was set to $3$ and the onium mass taken to give $s_0=0.1 \  GeV^2$. The value of the coupling, $\lambda = 23$, is fitted through the slope of the proton structure function $F_2$ in comparison to the DIS data. Phenomenological considerations on the saturation scale gives $\kappa=2.5$. These numerical values will be used in the following analysis.
The identification of the radial direction $z$ with the dipole size (inverse virtuality) and a comparison of the scaling of $F_2(x,Q^2)$ with $Q=1/z$ to the data confirms $D_\perp=3$ . The effective size of the proton and the position of the IR cutoff were taken as $z_p= 1.8 \ GeV^{-1}$, $z_0= 2 \ GeV^{-1}$.
While saturation effects become important at low $Q^2$ and small Bjorken $x$, both the confining result as well as the conformal limit for the scattering amplitude lead to a cross section that is comparable to the DIS data. \\ 

\section{Proton-proton scattering}

Diffractive proton-proton scattering at small momentum transfer unravels information about the transverse shape of the proton and the large $|t|$ behavior probes lengths scales of the typical string length, which in the confining background is of the order of $z_0$. We will fit the effective dipole size of the proton, $z_p$, and the position of the hard wall, $z_0$, to the data. All other numerical values remain the same as in \cite{Stoffers:2012zw}, see also section 2.\\

Instead of using diffractive eigenstates \cite{Good:1960ba}, \cite{Ryskin:2012ry}, perturbative \cite{Shoshi:2002in}, \cite{Wirbel:1985ji} or holographic light-front wave functions \cite{Brodsky:2006uqa,Brodsky:2006uq}, we will fit the data assuming the proton distribution is identified with the {\it wee-dipole} distribution, i.e. the proton is sharply peaked at some scale $1/z_p$, \cite{Brower:2010wf}. 
More explicitly, the square of the wave function will be approximated by a delta-function, $|\psi_p(u)|^2 =\mathcal{N}_p \  \delta(u-u_p)$. We treat the normalization constant $\mathcal{N}_p$ that carries the dipole distribution to the physical proton distribution as a parameter to be fitted to the data.

\subsection{Comparison to data: ISR}

\begin{figure}[t]
  \begin{center}
  \includegraphics[width=8cm]{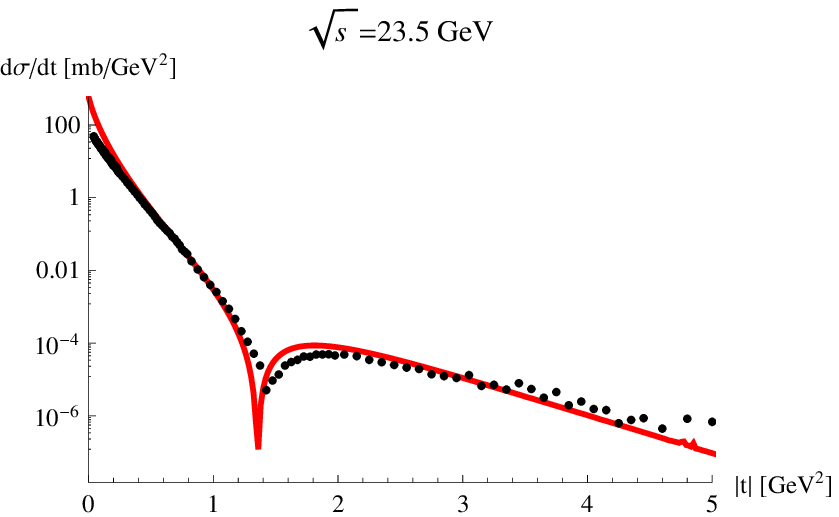}
  \includegraphics[width=8cm]{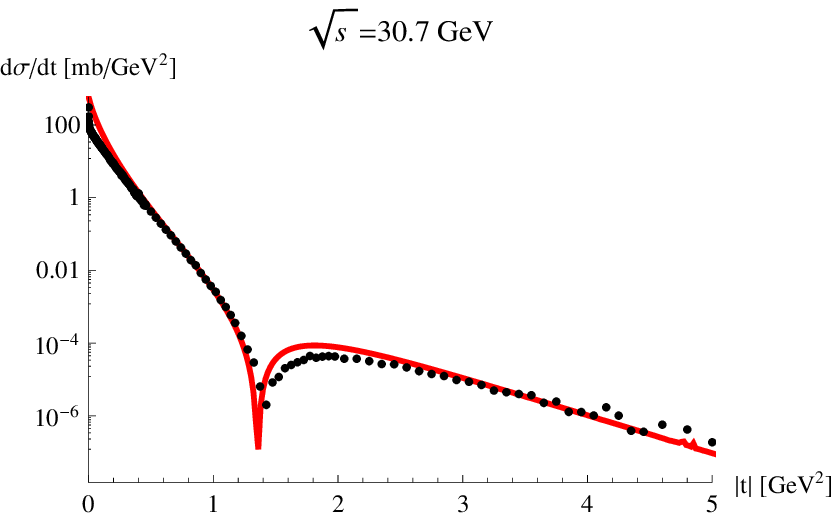}
  \includegraphics[width=8cm]{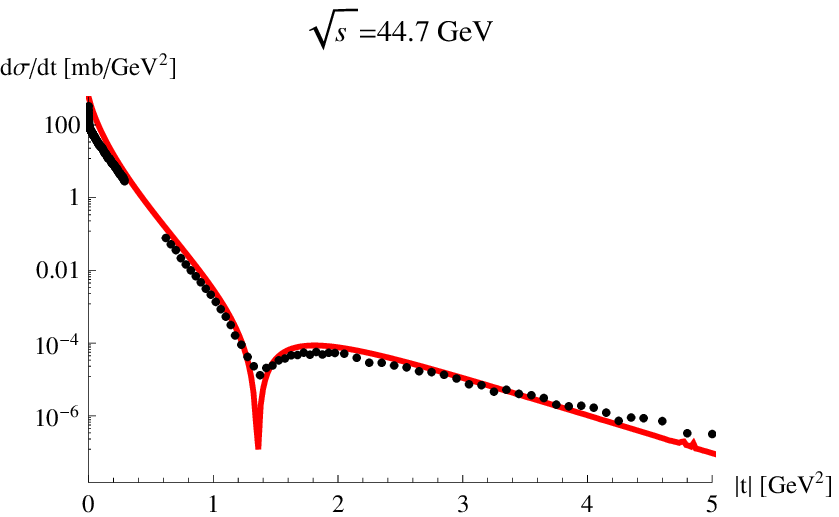}
  \includegraphics[width=8cm]{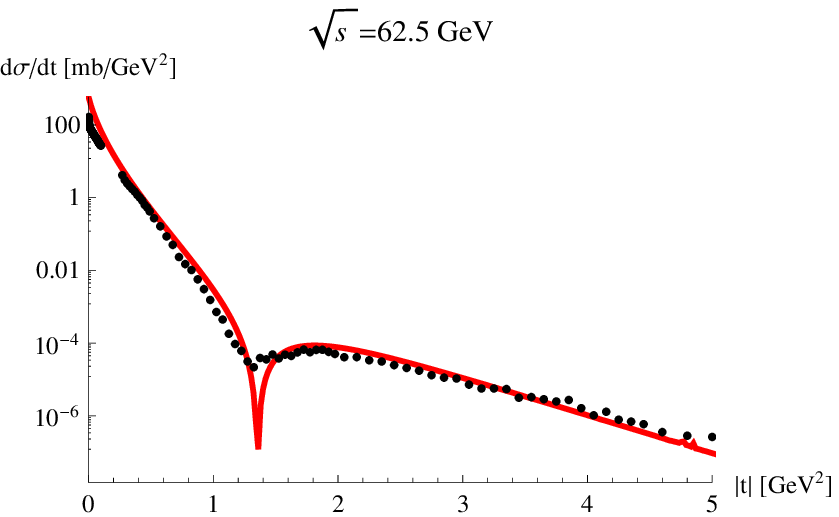}
  \caption{Differential $pp$ cross section. Dots: data from CERN ISR. Solid line: holographic result. See text.}
  \label{sigmadiff}
  \end{center}
\end{figure}

A comparison of the differential elastic $pp$ cross section, (\ref{dsigmadt}), to the CERN ISR data \cite{Amaldi:1979kd} is made by fitting the position of the dip and the slope of the shoulder region ($|t| > 1.5 \ GeV^2$). We use the full, unitary amplitude including higher order terms in the eikonal ${\bf WW}$. The importance of higher order terms in the eikonalized amplitude was also noted in \cite{Donnachie:2011aa}. A fit yields $z_0=2 \ GeV^{-1}$, $z_p=3.3 \ GeV^{-1}$ and  $\mathcal{N}_p=0.16$, see Figure \ref{sigmadiff}. To leading order, the position of the (first) dip is sensitive to the effective size of the scatterer and the energy of the scattering object and scales with $1/({\bf D} \chi z_p^2)$. The fit with $z_p > z_0$ is larger than the cutoff set by the hard-wall at $z_0$. This shortfall  is readily fixed by considering a smooth wall which is also more appropriate for describing hadron resonances \cite{Karch:2006pv}. The analysis of the dipole-dipole scattering amplitude in the smooth-wall background  will be discussed elsewhere.

At high momentum transfer ($|t| > 2 \ GeV^2$), the typical length scales probed are of the size of the fundamental string length, which is of the order of the IR cutoff. Thus, the slope of the shoulder region is fitted by primarily adjusting the value of the confinement scale $z_0$. The result for the cross section in the conformal limit $z_0 \rightarrow \infty$ does not yield a reasonable fit to the data. We
note that unlike perturbative QCD reasoning, \cite{Dremin:2012ke}, where the partons are resolved at large $|t|$ leading to a power-like decrease, the slope of the cross section at $|t|\ge 2 \ GeV^2$ is essentially not power-like in our holographic model.

\begin{figure}[t]
  \begin{center}
  \includegraphics[width=8cm]{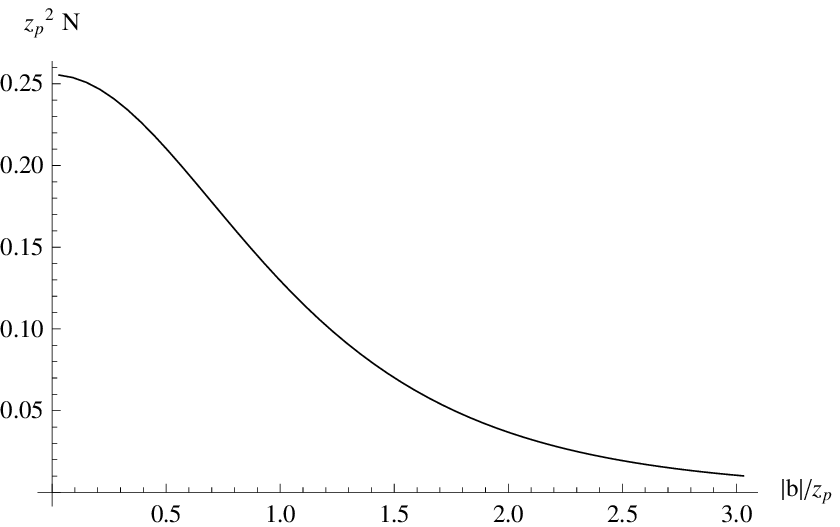}
  \includegraphics[width=8cm]{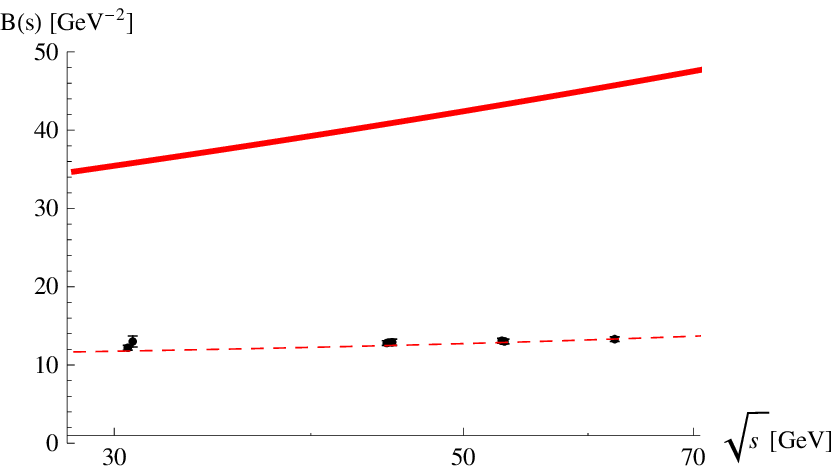}
  \caption{Left: Transverse distribution of the {\it wee-dipole} density ${\bf N}$, (\ref{dipoledensity}), with $\sqrt{s}=20 \ GeV$. Right:Experimental results for the slope parameter \cite{Amaldi:1971kt,Amaldi:1976yf} in comparison with the slope parameter, (\ref{B}), for $|t|=0 \ GeV^2$ (red, solid) and $|t|=1 \ GeV^2$ (red, dashed). See text.}
  \label{density}
  \end{center}
\end{figure}

At $|t| \sim 0 \  GeV^2$, the slope parameter $B(s,|t|)$ gives the mean square proton radius
\begin{eqnarray}
B(s,|t|=0)  \equiv  \left(\frac{d}{dt} \ln(\frac{d \sigma_{pp \rightarrow pp}}{dt}(s,t)) \right)\Big|_{t=0} = \frac{1}{2} \frac{\int d|{\bf b}|^2 \  |{\bf b}|^2 \left( 1-e^{\bf WW} \right) }{\int d|{\bf b}|^2 \left( 1-e^{{\bf WW}} \right) } = \frac{1}{2} <|{\bf b}|^2>  .\label{B}
\end{eqnarray}
The {\it wee-dipole} density ${\bf N}$ is peaked at $\frac{|{\bf b}|}{z_p}$ small, see Figure \ref{density},  and expanding the exponential to first order in $g_s^2$ gives
\begin{eqnarray}
B(s) \sim {\bf D} \chi (z_p^2+z_0^2) \ .
\end{eqnarray}
The radius of the proton is not only protortional to the effective {\it wee-dipole} size $z_p$ but also receives contributions from the IR cutoff.
At strong coupling, the diffusive nature of the eikonalized scattering amplitude is responsible for the scaling of the proton radius with the rapidity, $B(s)\sim {\bf D} \chi \sim \frac{1}{\sqrt{\lambda}} \ln \left(\frac{s}{s_0}\right)$. 

In the approach taken here, the transverse structure of the proton is modelled by a cloud of {\it wee-dipoles} surrounding a parent dipole. We can easily understand the scaling of the proton size with the coupling. As the coupling increases, the outer part of the cloud becomes more dilute and the proton shrinks.

Figure \ref{density} shows the slope parameter for $|t|=0 \ GeV^2$ and $|t|=1 \ GeV^2$. In our setup, the momentum distribution between the two constituents of each dipole is symmetric resulting in a small-size dipole, whereas asymmetric, large-size pairs dominate the small $|t|$ region, see e.g. \cite{Barone}. Thus, we suspect large-size dipoles to dominate the region $|t| \le 1 \ GeV^2$. The Coulomb contribution to the scattering amplitude can be neglected in the kinematic region $|t| > 0.01 \ GeV^2$ \cite{Amos:1985wx,Bernard:1987vq}.

\subsection{Comparison to data: LHC}

\begin{figure}[h]
  \begin{center}
  \includegraphics[width=12cm]{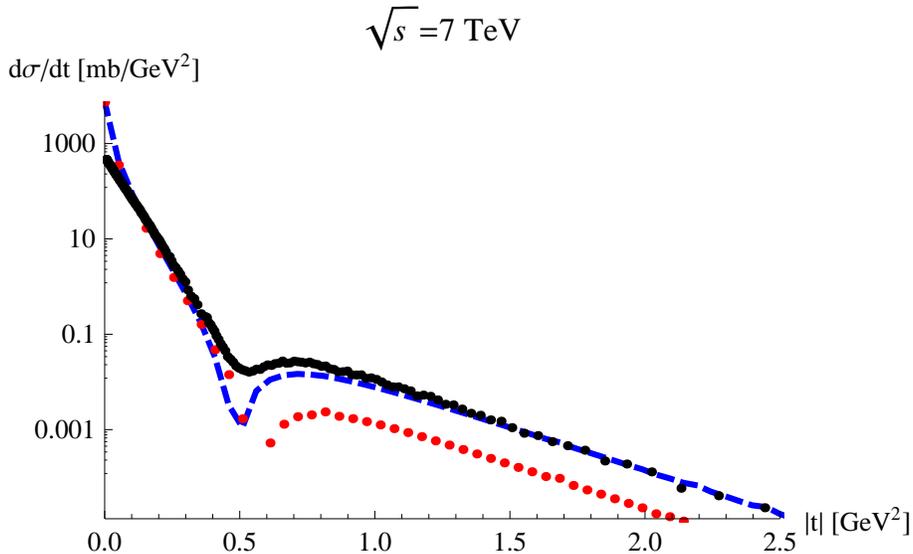}
  \caption{Differential $pp$ cross section. Black dots: data from the TOTEM experiment at LHC, \cite{TOTEM}. Dashed blue line and red dots: holographic result. See text.}
  \label{TOTEM}
  \end{center}
\end{figure}

Elastic $pp$ scattering at LHC energies of $\sqrt{s}= 7 \ TeV$, allows us to test the energy dependence of our model. With the numerical values fitted at energies $\sqrt{s} \sim 20-60 \ GeV$, the fit in Figure \ref{TOTEM} indicates a miss match in the energy dependence of the holographic model. In order to get a better fit to the LHC data, the parameters governing the overall strenght ($\kappa$), the position of the dip ($z_p$) and the slope of the shoulder ($z_0$) are adjusted. The fit (blue, dashed line) in Figure \ref{TOTEM} is obtained with $\kappa=3.75$, $z_p= 3.1 \ GeV^{-1}$, $z_0=1.5 \ GeV^{-1}$, while the fit (red, dotted) uses the values in section 3.1, $\kappa=2.5$, $z_p= 3.3 \ GeV^{-1}$, $z_0=2 \ GeV^{-1}$. This new parameter set for the LHC data is overall consistent with the set for the ISR data.

\section{Deeply virtual Compton scattering}

\begin{figure}[t]
  \begin{center}
  \includegraphics[width=8cm]{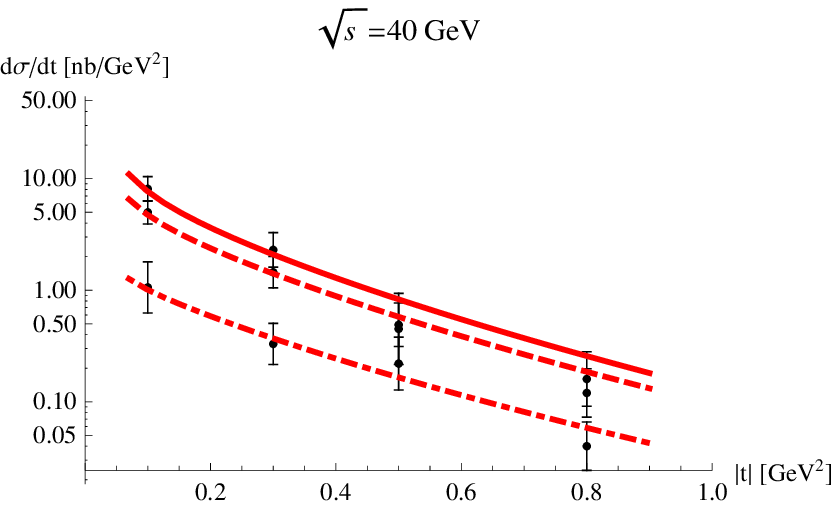}
  \includegraphics[width=8cm]{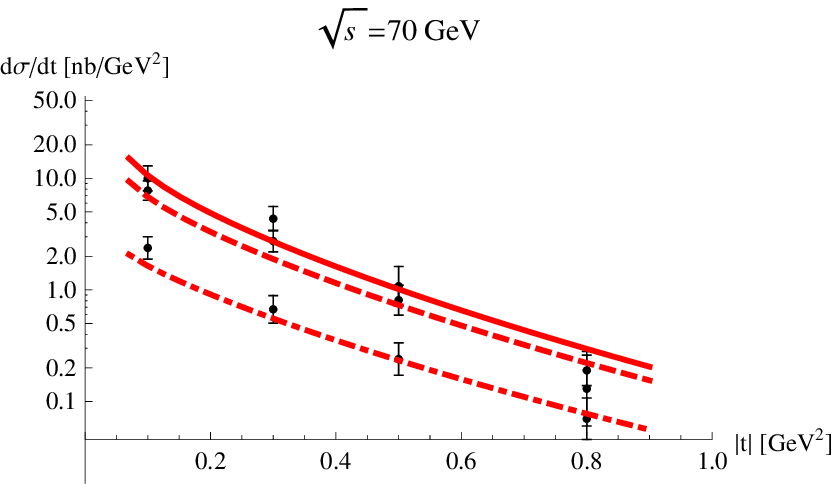}
  \includegraphics[width=8cm]{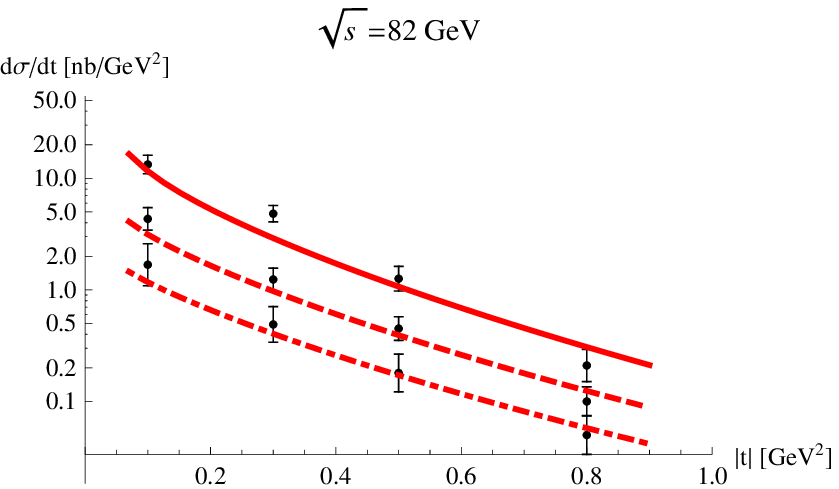}
  \includegraphics[width=8cm]{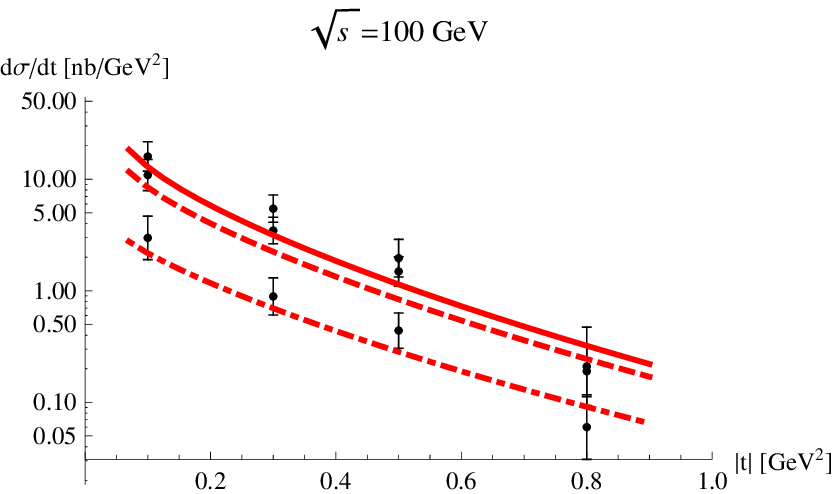}
  \caption{Holographic result for the differential DVCS cross section compared to the HERA data, \cite{Aaron:2007ab,Chekanov:2008vy,Aaron:2009ac}. $\sqrt{s}=82 \ GeV$: solid - $Q^2=8 \ GeV^2$, dashed - $Q^2=15.5 \ GeV^2$, dotdashed - $Q^2=25 \ GeV^2$. $\sqrt{s}=40, 70, 100 \ GeV$: solid - $Q^2=8 \ GeV^2$, dashed - $Q^2=10 \ GeV^2$, dotdashed - $Q^2=20 \ GeV^2$. See text. }
  \label{DVCS}
  \end{center}
\end{figure}

At high energies DVCS is dominated by pomeron exchange. In the rest frame of the proton, the virtual photon fluctuates into a quark-antiquark dipole that interacts with the proton. We will now use the dipole-dipole amplitude (\ref{WW}) to access the differential DVCS cross section $\frac{d \sigma_{\gamma^* p \rightarrow \gamma p}}{dt}$, (\ref{dsigmadtbessel}).
In section 3.1 we have refined the numerical values governing the transverse shape of the proton ($z_p$) and the IR cutoff scale ($z_0$) for the energy range of $\sqrt{s} \sim 20-60 \ GeV$. We will use these values to analyze the DVCS data in the range $\sqrt{s} \sim 40 - 100 \ GeV$. Now that all parameters of the holographic cross section are fixed, a comparison to the DVCS data serves as an additional test for our model.

The $\gamma^* \gamma$  overlap amplitude is approximated by a delta function, $|\psi_{\gamma^* \gamma} (u)|^2=\mathcal{N}_{\gamma* \gamma} \delta(u-u_{\gamma* \gamma})$, peaked at some finite virtuality $Q_{\gamma* \gamma} = 1/z_{\gamma* \gamma}$. The normalization constant is fitted to $\mathcal{N}_{\gamma* \gamma}=0.00016$. With the effective size of the proton, $z_p=3.3 \ GeV^{-1}$, and the position of the cutoff, $z_0=2 \ GeV^{-1}$, fixed, we compare our holographic result to the HERA data. Figure \ref{DVCS} illustrates an agreement of the cross section obtained from the holographic dipole-dipole scattering amplitude with the data. 
 
\section{Conclusions}

High energy hadronic scattering is dominated by pomeron exchange. Due to its non-perturbative nature at strong coupling, the holographic pomeron admits Gribov diffusion in curved space. Within the dipole-dipole scattering approach, the holographic description allows to access the saturation regime at small Bjorken $x$ and small momentum transfer.
The parameters of the model developed in \cite{Basar:2012jb} were fitted against DIS data in \cite{Stoffers:2012zw}. In order to refine the numerical values characterizing the proton shape and the IR cutoff, we have confronted the differential cross section with data on $pp$ scattering and DVCS. 

We have been able to get a reasonable fit to the $pp$ scattering data and obtained a refinement of the effective dipole size $z_p$ of the proton. The slope of the cross section in the region $|t|>2 \ GeV^2$ is sensitive to the IR cutoff scale, indicating the necessity of a confining background. However, the hard wall seems to be a too crude approximation for an IR cutoff. In order to fit the $pp$ data, we need $z_p \ge z_0$ suggesting that the smooth-wall background \cite{Karch:2006pv} is a more suitable setup. While the hard-wall construct allows for explicit and analytical results, the smooth-wall construct is likely numerical and will be addressed elsewhere.

The slope parameter $B(s,|t|)$ at small momentum transfer $t \sim 1 \ GeV^2$ agrees with the data. As is typical for diffusive processes, the mean square proton radius scales linear in rapidity. At strong coupling, the proton shrinks with increasing t'Hooft coupling.\\
Having fixed the parameters of the holographic model, an agreement with the DVCS data at small $|t|$ builds further confidence in the holographic approach to hadronic scattering.

\vskip0.5cm
{\bf Acknowledgements.}
A.S. would like to thank Frasher Loshaj for useful discussions.
This work was supported by the U.S. Department of Energy under Contract No. DE-FG-88ER40388.

\end{document}